\documentstyle[prl,aps,epsf]{revtex} 

\begin{document}

%
%

\twocolumn[\hsize\textwidth\columnwidth\hsize\csname%
@twocolumnfalse\endcsname%

\title{Phase Fluctuations and Vortex Lattice Melting in Triplet \\
Quasi-One-Dimensional Superconductors at High Magnetic Fields}

\author{C. D. Vaccarella and C. A. R. S\'a de Melo} 
\address{School of Physics, 
Georgia Institute of Technology, Atlanta, GA 30332} 

\date{\today}
\maketitle

\begin{abstract}
Assuming that the order parameter corresponds to 
an equal spin triplet pairing symmetry state, 
we calculate the effect of phase fluctuations in 
quasi-one-dimensional superconductors 
at high magnetic fields applied along the $y$ $(b^{\prime})$ axis.
We show that phase fluctuations can 
destroy the theoretically 
predicted triplet reentrant superconducting state, 
and that they are responsible for  
melting the magnetic field induced Josephson vortex lattice, above a
magnetic field dependent melting temperature $T_m$. 

\vspace{1pc}
\end{abstract}

]

%
%

The upper critical field $H_{c_2}$ of quasi-one-dimensional superconductors 
for perfectly aligned magnetic fields along the $y$ $(b^{\prime})$ axis 
has been calculated (at the mean field level) for both singlet and triplet 
states~\cite{lebed-86,lebed-87,dms-93}.
In the case of triplet pairing, these studies 
predicted a remarkable reentrant superconducting phase, 
for magnetic fields precisely aligned along
the $y$ $(b^{\prime})$ axis. 
New experiments performed in these systems
lead to the observation of unusual superconductivity,  
for instance, in quasi-one-dimensional superconductors (of the 
Bechgaard salts family with chemical formula ${\rm (TMTSF)_2 X}$, 
where ${\rm X = ClO_4, PF_6, ...}$) the experimental 
upper critical field exceeds substantially the Pauli paramagnetic
limit~\cite{lee-95,lee-97}. This indicates that high magnetic field 
superconductivity in these systems is most likely triplet. 
More recently, it has been argued 
by Lebed, Machida, and Ozaki~\cite{lebed-00} (LMO)
that the spin-orbit coupling must be strong in order to explain 
the observed experimental upper critical fields~\cite{lee-97}
and the absence of the Knight shift for fields parallel 
to the $y$ $(b^{\prime})$ axis.~\cite{lee-00}
However, estimates of the value of spin-orbit coupling 
are several orders of magnitude smaller~\cite{lee-98} 
than the values required to fit the critical
temperature of ${\rm (TMTSF)_2 PF_6}$~\cite{lebed-00,maki-89}, 
for instance, at low magnetic fields.
Very recently, the present authors~{\cite{sdm-00}} have analysed 
the angular dependence of $H_{c_2}$ for magnetic fields applied along the
$yz$ plane, where weak spin-orbit coupling and equal spin
triplet pairing were assumed. 
There, it was found that a strong suppression of the 
critical temperature $T_c (H)$ occurred for very small angular
deviations from the $y$ $(b^{\prime})$ axis. This result is in 
qualitative agreement with the rapid drop of the 
putative upper critical in ${\rm (TMTSF)_2 ClO_4}$~\cite{lee-98}.
However, all these theoretical 
works~\cite{lebed-86,lebed-87,dms-93,lebed-00,sdm-00},
have used a mean field approach, where the effects of 
fluctuations were completely ignored. 
Thus, the main point of the present paper is to discuss the 
effects of phase fluctuations on the field versus temperature
phase diagram of quasi-one-dimensional superconductors at high 
magnetic fields parallel to the $y$ $(b^{\prime})$ axis.
For that purpose, we study phase fluctuation effects
in the weak spin-orbit, equal spin triplet
state (ESTP) proposed by one us~\cite{sdm-96,sdm-98,sdm-99} 
as a possible candidate state for 
triplet superconductivity in the ${\rm (TMTSF)_2 X}$ family.
In particular, we show that phase fluctuations melt
the predicted magnetic field induced Josephson
vortex lattice at high magnetic fields~\cite{dms-93,sdm-96} 
at a melting temperature $T_m (H) < T_c (H)$. 
Furthermore, we show that the curvature of $T_m (H)$ is opposite
to that of $T_c (H)$ at high magnetic fields (range from 5 to 20 T), 
and that the superconducting state still exists at high magnetic
fields for temperatures $T < T_m (H)$.

We model quasi-one-dimensional systems 
via the energy relation
\begin{equation}
\label{eqn:dispersion-spin}
\varepsilon_{\alpha, \sigma}({\bf k}) =
\varepsilon_{\alpha} ({\bf k}) - \sigma \mu_B H,
\end{equation}
with the $\alpha$-branch dispersion 
\begin{equation}
\label{eqn:dispersion-0}
\varepsilon_{\alpha} ({\bf k}) = v_F (\alpha k_x - k_F) + 
t_y \cos(k_y b) + t_z \cos(k_z c),
\end{equation}
corresponding to an orthorombic crystal with lattice constants
$a,b$ and $c$~\cite{triclinic-00} along the $x,y$ and $z$ axis 
respectively. 
In addition, since $E_F = v_F k_F \gg  t_y  \gg t_z $,
the Fermi surface of such systems is not simply connected, 
being open both in 
the {\it xz} plane and in the {\it xy} plane. Furthermore, the 
electronic motion can be classified as {\it right}-going ($\alpha = +$) 
or {\it left}-going ($\alpha = -$). 

In this paper, we consider only magnetic field applied along the
$y$ $(b^{\prime})$ direction for simplicity, since
quantum confinement along the transverse $z$ direction occurs 
when $t_z/\omega_{c_z} \ll 1$, where $\omega_{c_z} = v_F G_z$
with  $G_z = |e| H_y c$.
Throughout the paper, we work in units where the Planck's constant and the 
speed of light are equal to one, i.e., $\hbar = c^* = 1$, 
and use the gauge 
${\bf A} = (H_y z, 0, 0)$, where $\alpha$ and $k_x$, $k_y$ are
still conserved quantities (good quantum numbers)
while $k_z$ is not.

In the presence of the magnetic field $H$, the non-interacting 
Hamiltonian is
\begin{equation}
H_0 ({\bf k} - e {\bf A}) = 
\varepsilon_{\alpha} ({\bf k} - e {\bf A}) - \sigma \mu_B H
\end{equation}
in the gauge ${\bf A} = (H_y z - H_z y, 0, 0)$.
The eigenfunctions of $H_0 ({\bf k} - e {\bf A})$ are 
\begin{equation}
\label{eqn:phiqn}
\Phi_{qn}({\bf r}) = \exp[i(k_{x}x + k_{y} y)] 
J_{N_z-n_z}\left({\alpha t_z \over \omega_{c_z} }\right), 
\end{equation}
where ${\bf r} = (x, y, z)$, $z = n_z c$ with associated quantum numbers 
$qn = \alpha,k_x,k_y,N_z,\sigma$ and 
eigenvalues
\begin{equation}
\label{eqn:eigenqn}
\epsilon_{qn} = 
\varepsilon_{\alpha, \sigma} ({\bf k}_{\rho}) 
+ \alpha N_z \omega_{c_z}.
\end{equation}
The function $J_{p}(u)$ 
is the Bessel function of integer order {\it p} and 
argument {\it u}, while now
\begin{equation}
\varepsilon_{\alpha, \sigma} ({\bf k}_{\rho}) = v_F(\alpha k_x - k_F) 
+ t_y \cos(k_y b) - \sigma \mu_B H 
\end{equation}
is a 2D dispersion. 
Notice that the electronic wavefunction in Eq.~(\ref{eqn:phiqn})
is confined along the $z$ direction when $t_z/\omega_{c_z} \ll 1$, 
thus limiting the electronic motion to a nearly two dimensional situation.
Furthermore, notice that the eigenspectrum in 
Eq.~(\ref{eqn:eigenqn}) involves 
many magnetic subbands labeled by quantum numbers $N_z$
and that the eigenvalue $\epsilon_{qn}$ is invariant under the 
quantum number transformation
$(\alpha,k_x, k_y, N_z, \sigma) \to (-\alpha,-k_x, -k_y, -N_z,\sigma)$,
for the same spin state $\sigma$.  
Since, these magnetic subbands are spin-split into 
{\it spin-up} and {\it spin-down} bands, Cooper pairs can be 
easily formed in a ESTP pairing state, involving electrons 
with quantum numbers $(\alpha,k_x, k_y, N_z, \sigma)$ and 
$(-\alpha,-k_x, -k_y, -N_z, \sigma)$, provided that the pairing 
interaction $\lambda_{\mu}$ conserves spin (which seems to be the case for the
${\rm (TMTSF)_2 X}$ family, except for very small spin-orbit and dipolar 
couplings)
Thus, in the ESTP state the order parameter vector 
\begin{equation}
\label{eqn:delta-vector}
{\vec \Delta } ({\bf r}) = \left[ \Delta_{+1} ({\bf r}), 0,
\Delta_{-1} ({\bf r}) \right] 
\end{equation}
has components only in the $m_s = +1$ $(\mu = \uparrow\uparrow)$ and 
$m_s = -1$ $(\mu = \downarrow\downarrow)$ channels, only. 

\begin{figure}
\begin{center}
\epsfxsize=7.0cm
\epsfysize=6.0cm
\epsffile{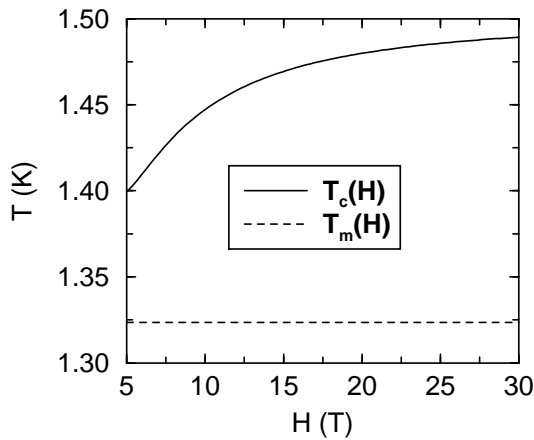}
\vskip 0.5cm
\caption{This figure shows the 
(mean field) critical temperature $T_c (H)$ and
the melting temperature $T_m (H)$, in the regime 
$t_z/\omega_{c_z} \ll 1$ for applied fields
${\bf H}$ precisely aligned with the $y$ $(b^\prime)$ axis.
The ratio $t_z/\omega_{c_z} \approx 1/3, 1/18$ for fields 
$H = 5, 30$ Teslas, respectively.
Temperatures are in Kelvin, and fields are in Teslas. 
The parameters used were 
$T_{c}(\infty) = 1.5 {\rm K}$, $t_y = 100{\rm K}$, $t_z = 5{\rm K}$ 
with lattice constants characteristic of Bechgaard salts.}
\label{fig:tctm}
\end{center}
\end{figure}

Now we turn our attention to the construction of the effective
free energy, where we assume that the interaction 
$\lambda_{\mu} = \lambda$ and the density of states 
${\cal N}_\mu = {\cal N}$ are independent of the
spin channel, which implies that 
$\Delta_{\mu} ({\bf r})  = \Delta ({\bf r})$.
Following the functional integral formulation of the ESTP discussed
by S\'a de Melo~\cite{sdm-96}, we derive the effective 
free energy functional ${\cal F}$ in the limit  
where the magnetic field is exactly pointing 
along the $y$ direction and where it nearly confines the 
electronic motion to a two dimensional regime ($t_z/\omega_{c_z} \ll 1$).
In this case, $\Delta ({\bf r}) = \Delta_{n} (x,y)$, where
$n$ now labels the planes $z = nc$, and the effective free energy 
takes the form
\begin{equation}
\label{eqn:free-energy}
{\cal F} =  \sum_n \int d{\bf r} \left(
{\cal F}_1 + {\cal F}_2 + {\cal F}_3 + {\cal F}_4 \right) + {\cal F}_{m}
\end{equation}
The first term is the local quadratic form
\begin{equation}
\label{eqn:f1}
{\cal F}_1 = a_1 \vert \Delta_n \vert^2,
\end{equation}
where the critical temperature can be derived from the coefficient
\begin{equation}
\label{eqn:a1}
a_1 = {\cal N} 
\left[ 
\ln \left({ T \over T_{c_{2d}} } \right) - 
\left( { t_z \over \omega_{c_z} } \right)^2 
\ln \left({\vert \omega_{c_z} \vert \gamma  \over \pi T} \right)
\right].
\end{equation}
The second term has the form 
\begin{equation}
\label{eqn:f2}
{\cal F}_2 = a_{2_x} \vert {\partial \Delta_n  / \partial x} \vert^2 
+ a_{2_y} \vert \left( { \partial / \partial y} - i 2 G_y x/ b \right) 
\Delta_n \vert^2,
\end{equation}
and corresponds to the spatial variation of $\Delta_n (x,y)$, with 
coefficients
$a_{2_x} =  {\cal N} 
\beta_t { C v_F^2 / T^2}$
and $a_{2_y} =  {\cal N} {C (t_y b)^2 / 2 T^2}$,
with $C = 7 \zeta (3)/ 16 \pi^2$. The third term corresponds to the
contribution of the magnetic field induced Josephson coupling
\begin{equation}
\label{eqn:f3}
{\cal F}_3 = a_3 \vert \Delta_{n+1} \exp \left( -i 2 G_zx/c \right) -
\Delta_n \vert^2, 
\end{equation}
with coefficient
$a_3 = {\cal N}  
\left( { t_z / \omega_{c_z} } \right)^2 
\ln \left({\vert \omega_{c_z} \vert \gamma  / 2\pi T} \right)
$.
While the fourth term corresponds to the non-local fourth order 
contribution
\begin{equation}
\label{eqn:f4}
{\cal F}_4 = a_{41} \vert \Delta_n \vert^4 + 
a_{42} \vert \Delta_n \vert^2 \vert \Delta_{n+1} \vert^2,
\end{equation}
where $a_{41}$ and $a_{42}$ are complicated functions of 
$t_y$, $t_z$, $\omega_{c_y}$ and $\omega_{c_z}$.
The last term 
$$
{\cal F}_m = \int d^3 {\bf r} { {\bf B}^2 \over 8 \pi}
$$
is just the magnetic energy. 
Using $\Delta_n = \vert \Delta_n\vert \exp (i \phi_{n})$, it is 
easy to show that the terms in Eqs. (\ref{eqn:f1}) and (\ref{eqn:f4})
do not contribute to the phase fluctuation 
free energy,
\begin{equation}
\label{eqn:fp}
{\cal F}_{p} =  \sum_n \int d {\bf r} 
\left( 
{\cal F}_{px} + {\cal F}_{py} + {\cal F}_{pz} \right)  + {\cal F}_{pm}.
\end{equation}
Using units where $\hbar = c^* = 1$, 
the first two contributions to the phase fluctuation free energy are
\begin{equation}
\label{eqn:fpmu}
{\cal F}_{p_\mu} = E_{\mu} 
\left\vert 
{\partial \over  \partial \mu} \phi_n - 
i 2e  A_{n_\mu} 
\right\vert^2, 
\end{equation}
where the characteristic energies in the $xy$ plane are
\begin{equation}
\label{eqn:emu}
E_{\mu} = \vert a_{2_\mu} \vert \vert \Delta_n \vert^2,
\end{equation}
with $\mu = x, y$. 
The third contribution is
\begin{equation}
\label{eqn:fpz}
{\cal F}_{p_\mu} = J_{z} 
\cos 
\left( 
\phi_{n+1} - \phi_n - 2G_z x -
2e {\bar A}_{n_z}
\right).
\end{equation}
where
${\bar A}_{n_z} = \int_{nc}^{(n+1)c} dz A_z /{\bar c}$, with the
Josephson energy term being
\begin{equation}
\label{eqn:ez}
E_z = {\bar c}^2 J_z =  a_3\vert \Delta_n \vert \vert \Delta_{n+1} \vert .
\end{equation}
The last contribution to the phase only fluctuation free energy is
\begin{equation}
\label{eqn:fpm}
{\cal F}_{p_m} = { 1 \over 8\pi} \int d^3 {\bf r} 
\left( {\bf \nabla \times A} \right)^2,
\end{equation}
where ${\bf A}$ is the fluctuation vector potential. 
The saddle point amplitude of the order parameter has the form 
\begin{equation}
\label{eqn:deltan}
\vert \Delta_n \vert
= \Delta_0 \left[ 1 + (t_z/{\omega_c})^2 \cos(2G_zx)\right],
\end{equation}
where the prefactor is
$\Delta_0 = \delta \left[ 1 + (t_z/\omega_{c_z} )^2 \right]$, 
(for $E_F \gg {\omega_{c_z}}$), with 
$
\delta = 
\sqrt{
T_c(H) \left[ T_c(H) - T \right]/ 2C}.
$
The saddle point phase $\phi_n^{(0)} = 2G_z n x - 
(8\pi a_3 \Delta_0^2/ H_y^2) n \sin(2G_z x)$. 
This corresponds to a rectangular Josephson 
vortex lattice with periodicity ${\it l}_x = \pi/G_z$ 
($G_z$ is the inverse magnetic length) and ${\it l}_z = {\bar c}$ 
(${\bar c}$ is the unit cell along the $z$ direction)~\cite{sdm-96} 
and holds a flux quantum $\phi_0$ inside the plaquette 
$({\it l}_x, {\it l}_z)$, i.e., 
$H {\it l}_x {\it l}_z = \phi_0$.~\cite{rectangular}
In what follows we use  
similar methods to those
developed by Horovitz~\cite{horovitz-a-92,horovitz-b-92} and 
Korshunov and Larkin~\cite{larkin-92} to study the vortex lattice
melting in layered superconductors. 
Writing down the phase of the order 
parameter as $\phi_n (x,y) = \phi_n^{(0)} + \chi_n (x,y)$, 
and integrating over the fluctuating vector potential we obtain 
the effective non-local sine-Gordon ``Hamiltonian''
\begin{equation}
\label{eqn:heff}
H = \int d{\bf r} \left( H_1 + H_2 \right)
\end{equation}
\begin{equation}
\label{eqn:h1}
H_1 =
\sum_\mu 
\left( 
E_\mu \sum_{n, n^\prime} 
\gamma_\mu (n, n^{\prime})
{\partial \chi_n \over \partial \mu}
{\partial \chi_{n^{\prime}} \over \partial \mu}
\right)
\end{equation}
\begin{equation}
\label{eqn:h2}
H_2 = 
- \Gamma \sum_n \cos(\chi_{n+1} - 2\chi_n + \chi_{n-1}),
\end{equation}
where the co-sinusoidal coupling constant is 
\begin{equation}
\label{eqn:gamma}
\Gamma = J_z^2/4\sqrt{E_x E_y}\sqrt{{\tilde\gamma}_x(\pi)
{\tilde\gamma}_y(\pi)}h_z^2
\end{equation}
with the function
\begin{equation}
\label{eqn:tgamma}
{\tilde\gamma}_{\mu}(q) = {\lambda (q) \over \lambda (q) + 
{16\pi^3 E_{\mu} d/ \phi_0^2} }
\end{equation}
being the 
discrete Fourier transform of $\gamma (n, n^{\prime})$, 
where $\lambda (q) = [1 - \cos(q)]$. 
When $\Gamma \to 0$,
the effective Hamiltonian given in Eq.~(\ref{eqn:heff})
reduces to a layered anisotropic XY model, where phase fluctuations 
between layers $n$ and $n^{\prime}$ are still coupled via the
function $\gamma_{\mu} (n, n^{\prime})$. 
Scaling the integration variables 
$x \to {\tilde x}\sqrt{E_x \gamma_x}$ and
$y \to {\tilde y}\sqrt{E_y \gamma_y}$, 
and defining the charges 
$q_{n_i} = \oint_{\partial{\bf r}_i} d \chi_n ({\bf r}) /2\pi$, 
which correspond to the vorticities 
at position ${\bf r}_i$ in the $n$-th plane,
leads to a partition function identical to the anisotropic 
quasi-two-dimensional Coulomb gas, 
when an expansion in powers of the very small parameter $\Gamma$ 
and Gaussian integration over $\chi_n$ 
are performed~\cite{larkin-92,kosterlitz-74}, i.e.,
\begin{equation}
\label{eqn:partition}
Z = \sum_{m} 
\Big\lbrace 
\prod_{i = 0}^{2m} 
\left( 
\int d {\bf r}_i \sum_{n_i} 
\right)  
\Big\rbrace
\left( 
{\Gamma \over 2 T} 
\right)^{2m} \exp (A), 
\end{equation}
where the exponent in the partition function is 
\begin{equation}
\label{eqn:A}
A =  
- {1 \over 2} \sum_{i,j}^{2m} q_{n_i} M ({\bf r}_{ij}, n_{ij}) q_{n_j}
\end{equation}
where the charges $q_{n_i} = +1$ for $i = 1, ...., m$ and
$q_{n_i} = -1$ for $i = m + 1, ...., 2m$. In addition, 
${\bf r}_{ij} = {\bf r}_j - {\bf r}_i$ and $n_{ij} = n_j - n_i$, 
and the correlation function
$$
M({\bf r}_{ij}, n_{ij}) = 
\int {d^2 k \over (2\pi)^2} {dq\over 2\pi} 
{\tilde M} ({\bf k}, q) 
\exp\left[ i ({\bf k}\cdot {\bf r}_{ij} + n_{ij} q ) \right]
$$
has as Fourier transform the function
${\tilde M} ({\bf k}, q) = 
{T /\sqrt{E_x E_y} k^2 {\tilde\gamma} (q)},
$
with ${\tilde\gamma} (q) = \sqrt{ {\tilde\gamma}_x (q){\tilde\gamma}_y (q) }$.
Upon integration over $k$ and proper core 
regularization, the relevant part of the interaction term becomes
\begin{equation}
\label{eqn:amij}
M({\bf r}_{ij}, n_{ij}) \approx 
{ \alpha(n_{ij})T \over 2\pi \sqrt{E_x E_y} } 
\ln(r_{ij}/a_0)
\end{equation}
where $\alpha (n_{ij}) = 
\int dq \exp(i n_{ij}q)/[2\pi {\tilde \gamma} (q)]$.
Notice that the charges interact logarithmically whether they are in 
the same layer or not; 
however, the strength of the logarithmic interaction is larger
when they are in the same layer. 
This means that
in the limit where $\Gamma \to 0$ there is a phase transition 
which belongs to the same universality class of the 
Kosterlitz-Thouless-Berezinskii transition~\cite{kt-72,berezinskii-72}, 
and corresponds, in the present problem,
to the melting of the magnetic field induced 
rectangular Josephson vortex lattice.
The transition here occurs when 
\begin{equation}
\label{eqn:tm-exp}
T_m  = {\pi \sqrt{E_x (T_m) E_y (T_m )} 
\over 1 + \sqrt{E_x (T_m) E_y (T_m)} 
16 \pi^3 {\bar c}/\phi_0^2}.
\end{equation}
Notice that the right hand side of Eq.~\ref{eqn:tm-exp} 
is also dependent on temperature via
$E_{\mu} (T)$. Using the expressions for 
$E_{\mu} (T)$ defined in Eq.~(\ref{eqn:emu}) and solving equation
Eq.~(\ref{eqn:tm-exp}) for the melting temperature at infinite field
results in
\begin{equation}
\label{eqn:tm}
T_m (\infty) = { T_c (\infty) \over 1 + \eta }
\end{equation}
where $\eta = 2\pi \sqrt{2} T_c (\infty)/t_y$. 
Since $\eta > 0$, the melting 
temperature $T_m (\infty)$ is smaller than the saddle point (mean field)
critical temperature $T_c (\infty)$, thus indicating that classical phase
fluctuations reduce the transition temperature from $T_c (\infty)$ to
$T_m (\infty)$. However, this reduction is small for Bechgaard salts
since $\eta \ll 1$.
When the condition $H \to \infty$ is relaxed the first
order correction to $T_m$ can be computed using a perturbative renormalization 
group method~\cite{horovitz-a-92,horovitz-b-92,kosterlitz-74}
since $E_z$ is much smaller than $T_c (\infty)$ and $T_m (\infty)$. 
This standard procedure leads to 
\begin{equation}
\label{eqn:tm-ren}
T_m (H) = T_m (\infty) 
\left[ 
1 + \epsilon \left( {\ell_0^2 \Gamma \over 2 T_m (\infty)}\right)^2,
\right] 
\end{equation}
where $\epsilon = \pi/2$, $\ell_0 = \pi/\omega_{c_z}$ and 
the magnetic field dependend coefficient 
$$
\Gamma = {\cal N} T_c^2 (\infty) 
\left( { {\pi^2}\sqrt{2} v_F \over 7 \zeta (3) t_y b} \right)
\left( { \Delta_0^2 t_z^4 \over \omega_{c_z}^6 } \right)
\ln^2 \left( { \gamma \vert \omega_{c_z} \vert \over 2 \pi T_c (\infty) }
\right).
$$
Notice the explicit magnetic field dependence of 
$\Gamma \sim H^{-6}\ln^2 (H)$, such that the correction to $T_m (\infty)$
increases with decreasing magnetic field as $T_m (H) - T_m (\infty) 
\sim H^{-16} \ln^4 (H)$ in the high field regime 
where $t_z/\omega_{c_z} \ll 1$.
This behavior has the following interpretation: a reduction of 
the magnetic field increases the magnetic field induced 
Josephson coupling $E_z$ and the system starts to become
less two-dimensional, so that phase fluctuations become less efficient
and $T_m (H)$ increases with decreasing field. However, the increase in 
$T_m (H)$ with decreasing field is very slow, and it looks quite flat
when $T_c (H)$ and $T_m (H)$ are plotted 
in the same scale~(See Fig.~\ref{fig:tctm}).
This is in sharp contrast with
what happens with the saddle point (mean field) critical temperature, 
which at high magnetic fields decreases from $T_c (\infty)$ 
as the magnetic field is lowered. The difference in behavior between
$T_c (H)$ and $T_m (H)$ is illustrated in Fig.~\ref{fig:tctm}.
\vskip -0.5cm

In summary, 
we have assumed an ESTP state as a plausible candidate for triplet
superconductivity in quasi-one-dimensional 
systems~\cite{sdm-96,sdm-98,sdm-99},
and we have presented analytical results for the effects of phase 
fluctuations in the magnetic field versus temperature phase 
diagram of quasi-one-dimenional superconductors. We discussed 
for simplicity just the case of perfect alignment with the
$y$ $(b^{\prime})$ direction, and showed that phase fluctuations
destroy the reentrant superconducting phase when 
$t_z / \omega_{c_z} \ll 1$. This loss of phase coherence corresponds
to the melting of the magnetic field induced Josephson vortex lattice
above the melting temperature $T_m (H)$. At very large fields $T_m (H)$
is only slightly smaller than the mean field critical temperature $T_c (H)$,
however $T_m (H)$ has the opposite curvature of $T_c (H)$. This means that 
$T_m (H)$ decreases with increasing field while $T_c (H)$ 
increases with increasing field, provided that 
the condition $t_z/\omega_{c_z} \ll 1$ is satisfied. Furthermore,
$T_m (H) - T_m (\infty) \sim H^{-16} \ln^4 (H)$ in this high field regime,
and $T_m (H)$ looks quite flat when plotted in the same scale as $T_c (H)$.
Finally, it is important to emphasize that we 
have discussed here the effects of classical phase fluctuations 
for perfect alignment along the ${b}^{\prime}$ 
direction only.~\cite{footnote} 

We would like to thank the Georgia Institute of Technology, 
NSF (Grant No. DMR-9803111) and 
NATO (Grant No. CRG-972261) for financial support, 
and one of us (C. A. R. S. d. M.) would like to thank also 
the Aspen Center for Physics for their hospitality.


\begin{references}
\bibitem{triclinic-00} 
The Bechgaard salt family ${\rm (TMTSF)_2 X}$ is truly
triclinic, however, many of the qualitative results discussed in this
paper can be easily generalized to the triclinic case.
\bibitem{rectangular} The Josephson vortex lattice (calculated
in mean field theory) is rectangular, instead of triangular as
considered in Ref.~\cite{dms-93}. The assumed orthorombic symmetry
of the underlying lattice imposes serious restrictions on the boundary
conditions for the spatial profiles of the magnetic field, and thus 
frustrates the triangular solution in favor of the rectangular.
\bibitem{footnote} We are aware that the 
effects of quantum (temporal) fluctuations, 
amplitude fluctuations, and the direction of the applied magnetic field 
can further reduce the melting temperature $T_m (H)$. 
These effects will be investigated at a later date.
\end{references}
\end{document}